\documentclass[10pt,twocolumn,letterpaper]{article}

\usepackage{times}
\usepackage{epsfig}
\usepackage{graphicx}
\usepackage{amsmath}
\usepackage{amssymb}
\usepackage{multirow}
\usepackage[symbol]{footmisc}
\usepackage{algorithm2e}
\usepackage{cite}
\usepackage{authblk}

\usepackage[pagebackref=false,breaklinks=true,letterpaper=true,colorlinks,bookmarks=false]{hyperref}

\DeclareMathOperator*{\argmax}{arg\,max}

\title{Graph Neural Network for Hamiltonian-Based Material Property Prediction}

\author[1]{Hexin Bai}
\affil[1]{Temple University, hexin.bai@temple.edu}
\author[2]{Peng Chu}
\affil[2]{Temple University, peng.chu@temple.edu}
\author[3]{Jeng-Yuan Tsai}
\affil[3]{Temple University, jeng-yuan.tsai@temple.edu}
\author[4]{Nathan Wilson}
\affil[4]{Texas A\&M University, wilsonnater@tamu.edu}
\author[5]{Xiaofeng Qian}
\affil[5]{Texas A\&M University, feng@tamu.edu, corresponding author}
\author[6]{Qimin Yan}
\affil[6]{Temple University, qiminyan@temple.edu, corresponding author}
\author[7]{Haibin Ling}
\affil[7]{Stony Brook University, hling@cs.stonybrook.edu, corresponding author}

\begin{document}
\maketitle
\section{abstract}
Development of next-generation electronic devices for applications call for the discovery of quantum materials hosting novel electronic, magnetic, and topological properties. Traditional electronic structure methods require expensive computation time and memory consumption, thus a fast and accurate prediction model is desired with increasing importance. Representing the interactions among atomic orbitals in any material, a material Hamiltonian provides all the essential elements that control the structure-property correlations in inorganic compounds. Effective learning of material Hamiltonian by developing machine learning methodologies therefore offers a transformative approach to accelerate the discovery and design of quantum materials. With this motivation, we present and compare several different graph convolution networks that are able to predict the band gap for inorganic materials. The models are developed to incorporate two different features: the information of each orbital itself and the interaction between each other. The information of each orbital includes the name, relative coordinates with respect to the center of super cell and the atom number, while the interaction between orbitals are represented by the Hamiltonian matrix. The results show that our model can get a promising prediction accuracy with cross-validation. 

\section{Introduction}

A key bottleneck limiting the applications of current artificial intelligent (AI) techniques, such as convolutional neural networks (CNN)~\cite{lecun1998gradient}, for quantum material discovery is the great diversity in condensed morphology of materials. To accelerate the discovery of functional inorganic materials using a data-driven approach, effective learning methods to establish the connections among materials are essential. In recent years, successful CNN-based methods have been developed for signals, e.g. images~\cite{krizhevsky2012imagenet}, where the local neighborhood for convolution operation is unique and fixed. CNNs have shown promising achievements on real-space wavefunction-based analysis in small molecules and orthogonal systems. However, for nonorthogonal grids and large molecules, CNN-based methods are found to be insufficient for the proper modeling since the locality in infinite or nonorthogonal systems differs substantially from that in finite orthogonal systems. Moreover, the diversity in grid and scale also prohibits generalization of models trained on one system to another.

Containing comprehensive and complete information of material systems, a Hamiltonian matrix governs the fundamental physics and electronic structure related complex material behaviors such as electronic, magnetic, and topological properties. It is a subject of continuous interest to compute Hamiltonian matrices directly from first-principles calculations in a tight-binding setup . Accurate tight-binding Hamiltonian can be extracted from first-principles calculations using the quasi-atomic minimal-basis-set orbitals~\cite{qian2008qo} readily available in the Wannier90 code~\cite{marzari1997mlwf, mostofi2014wannier}. These first-principles tight-binding Hamiltonian matrices provide a direct and transparent picture of chemical bonding, revealing physical quantities such as local densities of states and bond orders. Furthermore, the development of the Hamiltonian matrix enables us to perform tight-binding electronic structure calculations for large systems which requires much less computational time and memory than full electronic structure calculations. As a fundamental set of material information, Hamiltonian matrix can serve as a novel input to completely represent inorganic crystalline systems and be incorporated into a network-based machine learning framework to provide the fundamental understanding of orbital-based quantum mechanical interactions and accelerate quantum material discovery.

The basis of atoms and mutual interactions in quantum materials naturally encourages using the more general graph representation for materials. In particular, locality in quantum material systems can be defined by graph nodes independently from their actual grid and scale in real-space with the interaction between elements represented by graph edges. The recent emerging graph convolutional networks~\cite{battaglia2018relational,bronstein2017geometric, kipf2016semi, henaff2015deep} enable a flexible modeling of various grid and atomic structures. Furthermore, symmetry information of the condensed system can be easily integrated into the spectral formulation of graph convolution, which is essential for determining the physical property of a solid-state quantum material.

In this paper, we explore two different Graph Convolutional Network (GCN) architectures to predict the material properties of the quantum materials using their Hamiltonian representation. The Message Passing Graph Network conducts the information aggregation of neighboring nodes in the graph in the real space. On the other hand, by transferring the graph Laplacian into Fourier domain, Chebyshev Convolution leverages the Chebyshev polynomial to accelerate the convolutional operation in the spectral domain. The two methods are tested on our collected dataset to predict the band gap. Compared with traditional hand-crafted feature-based methods, the GCN-based ones generate clearly better performance on this binary classification task. 

\section{Related Work}

Machine learning techniques provide a novel opportunity to significantly reduce computational costs and speed up the pace of materials discovery by utilizing data-driven paradigms.\cite{Rajan2005,Ghiringhelli2015,Pilania2016} Instead of numerically solving complex systems with quantum interactions, physical quantities are statistically estimated based on a reference set of known solutions. Machine learning, especially supervised learning, has been effectively applied to materials property predictions, including phase stability (for both molecules and crystals),\cite{Long2007} crystal structure,\cite{Curtarolo2003,Hautier2010} electronic structure,\cite{Pilania2016,Lee2016} molecule atomization energies,\cite{Hansen2013} effective potential for molecule dynamics simulations,\cite{Behler2011,Bartok2010} and energy functional for density functional theory based simulations.\cite{Snyder2012} Neural network has been applied to inorganic crystal systems with limited applications. Band gaps of given classes of inorganic compounds have been predicted using deep neural networks.\cite{Dong2019,Shi2019} Recently, machine learning has been applied to 3D volume data such as electron charge densities to predict electronic properties.\cite{Chandrasekaran2019}

Recently, there has been great advance in the applying Graph neural network (GNN) for pattern recognition and data mining. Zhou et al. gave a comprehensive literature review in \cite{zhou2018graph} which covered recent methods and applications of GNN in modeling physics system, learning molecular fingerprints, and predicting protein interface. In~\cite{battaglia2018relational}, Battaglia et al. applied relational inductive biases in deep learning and presented a highly generalized framework for GNN. With the increasing number of GNN architectures appeared recently, \cite{dwivedi2020benchmarking} proposed a reproducible benchmark framework for GNN to gauge their effectiveness and compare with peer methods.  


\section{Method}

\subsection{Problem Formulation}

In materials sciences, the material band gap is an important property governing whether the material is metal or non-metal. In this study, we aim to use GCN to predict the band gap given the Hamiltonian of the material. Band gap is described by a non-negative real number, $E_g \in \mathbb{R}$ and $E_g \ge 0$. To simplify the problem, threshold is applied to sort $E_g$ into two categories $c \in \{0, 1\}$ to represent the metal and non-metal classes as the learning target, respectively. Finally, the learning problem is defined as:
\begin{equation}
    \hat{c} = \argmax_{\theta}f\big(c; \theta, x \big),
\end{equation}
where $\hat{c}$ is the prediction, $x$ the input representation of Hamiltonian, and $f(\cdot)$ the function with trainable parameters $\theta$ to be determined in this study.


\subsection{Hamiltonian Matrix}
\begin{figure}
	\includegraphics[width=\linewidth]{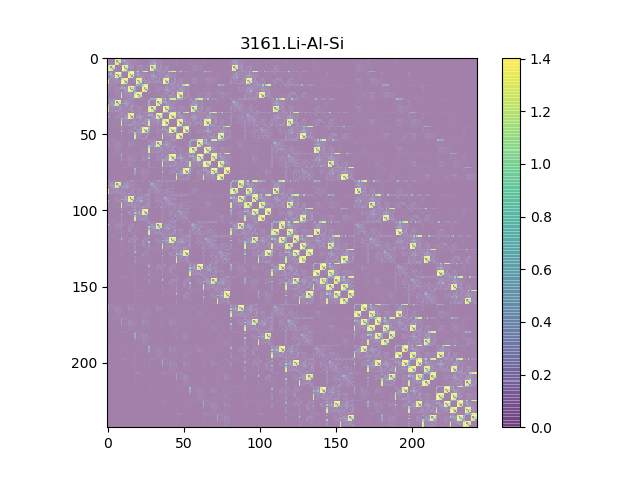}
	\caption{The Hamiltonian matrix of Li-Al-Si}
	\label{fig:Hamivisual}
\end{figure}

We will focus on the 2D Hamiltonian matrix that governs the fundamental physics of the quantum materials. Specially, Hamiltonian of a physical system contains all operators corresponding to the kinetic and potential energies. Let the Hamiltonian operator $\hat{H}$ for an $N$-particle condensed matter system be
 \begin{equation}
   \hat{H} = \sum_{n=1}^N \hat{T}_n + V,
 \end{equation}
where $\hat{T}_n $ indicates the kinetic energy operators for each particle, and $V=V(\mathbf{r}_1, \mathbf{r}_2, \dots, \mathbf{r}_N, t)$ is the potential energy between particles.  
To facilitate the calculation, the Hamiltonian operator can be represented as a 2D numerical matrix. In detail, the representation of an operator can be obtained through the integral with the basis of a Hilbert space. In the condense matter system, the wavefunctions $\{\varphi_i\}$ of orbitals from all atoms in the material system can form a Hilbert space. Therefore, the element in the matrix representation of Hamiltonian $H_{ij}$ can be calculated from
 \begin{equation}
     H_{ij} = \int \varphi_i \hat{H} \varphi_j d^3 r,
\label{eq:h-matrix}
 \end{equation}
resulting in an $M \times M$ Hamiltonian matrix $H=\{H_{ij}\}$ with $M$ as the total number of orbitals. In this paper, the Hamiltonian is computed from a super cell of $7\times 7 \times 7=343$ unit cells. Each cell contains around 3 atoms with certain number of orbitals. In Fig.~\ref{fig:Hamivisual}, we visualize the Hamiltonian matrix for a sample in the Li-Al-Si material system with $M=432$.
Note that in order to keep consistency with our experiment, we consider only the center $3\times 3 \times 3=27$ for computation cost. In the example of Fig.~\ref{fig:Hamivisual}, with $9$ orbitals, we got a square matrix of dimension $27\times 9 = 243$ filled with the real value which is quite closed to the norm given the imaginary part is very small.



\subsection{Graph Representation of Hamiltonian}
\label{feature}

Hamiltonian contains more concentrated information and has much lower input dimension than wavefunctions or charge density when used for physical property prediction. Thus, directly using Hamiltonian for prediction may reduce the model complexity and further relieve the demand of big data for model training. However, due to different atom composition in each condensed matter system, the size of Hamiltonian varies greatly. To handle the diversity of input dimension, we propose using a graph to store the Hamiltonian matrix for this learning task.

A weighted undirected graph can be constructed from the Hamiltonian matrix to encode all interactions and symmetric information of the quantum system. As indicated in Eq.~\ref{eq:h-matrix}, the matrix element $H_{ij}$ represents the interaction intensity between the $i$-th orbital and the $j$-th orbital in the condensed matter system. Then, if the orbitals are represented by $M$ vertices in a graph, such as $\mathcal{V}=\{\mathbf{v}_i\}_{i=1:M}$, naturally, the interaction between orbitals can be described by the edges in the graph. Noting that, according to Eq.~\ref{eq:h-matrix}, interactions exist for each pair of orbitals in the system, which results in a complete graph. In this work, we only consider the interaction stronger than a threshold $\tau_h$ to reduce the computational burden. Finally, a graph $\mathcal{G}=(c, \mathcal{V},\mathcal{E})$, with $\mathcal{E} = \{(\mathbf{e}_{r_k s_k}, r_k, s_k)\}_{k=1:N}$ for $N$ edges, is built for each Hamiltonian matrix. Specifically, let $r_k = i$ and $s_k = j$ indicate the two nodes associated with the $k$-th edge; an edge $(\mathbf{e}_{r_k s_k}, r_k, s_k)$ captures the interaction between $i$-th and $j$-th orbitals; its weight $\mathbf{e}_{ij}  = real(H_{ij})$  stands for the real part of complex number $H_{ij}$ in Hamiltonian matrix. In practice the scale of the imaginary part is negligible comparing with
the real part. Hence the real part of the complex number is a close approximation to the modulus of the complex number. 


\begin{table*}
	\caption{Feature component used in the GCN methods.}
	\label{tab:feat}
	\centering
\begin{tabular}{c|ccc}
\hline\hline
                         & Feature& Type & Description   \\
                         \hline
\multirow{3}{*}{Node}    & Atomic Number & 1 Integer  &  Number of protons    \\
                         & Atom Coordinates& 3 Real  & 3D location in the unit cell   \\
                         & Orbit Type  & $1\times16$ One hot  & $s, p, d, \dots $ \\
\hline
\multirow{1}{*}{Edge} & Interaction & 1 Real & Real part of $H_{ij}$\\
\hline\hline
\end{tabular}
\end{table*}

\begin{figure}[h]
	\includegraphics[width=\linewidth]{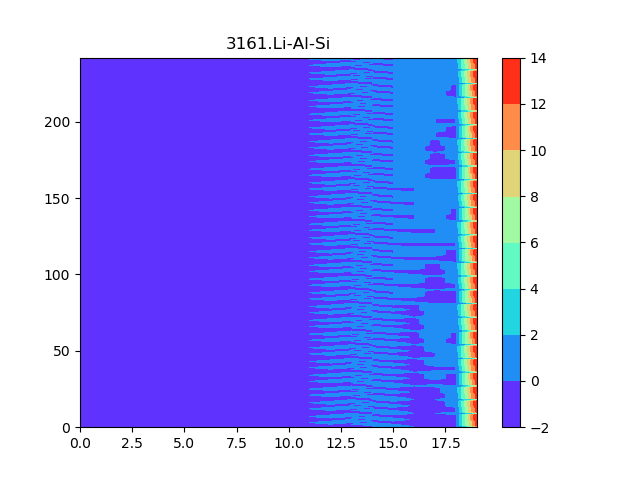}
	\caption{Visualization of node features for Li-Al-Si. Each row of the matrix represents a node feature in the $3\times 3 \times 3$ unit cells. Each node feature contains atomic number, coordinates and orbit type with detail in Tab.~\ref{tab:feat}}
	\label{fig:nodevisual}
\end{figure}

For each node inside the graph, we have the following node feature vector representation. In Tab.~\ref{tab:feat}, a summary of feature components for the node and edges is listed. Specifically, for each node, features encode the information for each orbital, which fuse both atom and orbital information. The collection of node features for all nodes in $3\times3\times 3$ unit cells of a Li-Al-Si sample is visualized in Fig.~\ref{fig:nodevisual}.  Edges, as we discussed above, represent the interaction between orbits, thus directly use the real part $H_{ij}$ as feature. 

With the graph represented Hamiltonian, we investigate two types of GCNs to explore and learn the information in the Hamiltonian for the band gap prediction.

\subsection{Message Passing Graph Network}

A single graph convolutional block is composed of update functions $\phi$ and aggregation functions $\rho$, such as 
\begin{equation} \label{GNblock}
\begin{split}
\mathbf{x}_{ik} &= \psi(\mathbf{e}_{ik},\mathbf{v}_{k}), \ k\in \mathcal{N}(i)\\
\mathbf{\bar{x}}_{i} &= \rho_\mathrm{local}\Big(\bigcup_{k\in \mathcal{N}(i)} \mathbf{x}_{ik}\Big)
 \\
\mathbf{v}_i &= \phi^{\mathbf{v}}(\mathbf{\bar{x}}_{i}, \mathbf{v}_i)\\
\mathbf{\bar{v}} &= \rho_{\mathrm{global}}\Big(\bigcup_{i}\mathbf{v}_i\Big)\\
\hat{c} &= \phi^c(\mathbf{\bar{v}})
\end{split}
\end{equation}
where $\mathcal{N}(i)$ stands for the neighborhood of node $i$. The order-invariant aggregation functions $\rho$ must provide the same output no matter what permutation of input is. $\rho_\mathrm{local}$ aggregates information from all the edges connecting to the node $i$ while $\rho_\mathrm{global}$ summarizes the information globally. The update functions  $\phi^{\mathbf{v}}$ and $\phi^c$ update the node and global attribute respectively. The  $\mathbf{x}_{ik}$, updated by function $\psi $, is a learned vector representation for each edge while the original edge attribute is kept unchanged to provide further critical information.
Note that this learned edge feature vector is also updated due to refreshing the node representation.

This general framework can be implemented with different flexible variants. In~\cite{gilmer2017neural}, the Message Passing Neural Network is proposed to allow long range interactions between nodes in the graph for molecular properties prediction. A modified version under this general framework is implemented using Eq.~\ref{GNblock} as explained in detail in Alg.~\ref{MPNN_algo}. 

\begin{algorithm}\label{MPNN_algo}
\SetAlgoLined
\SetKwProg{Fn}{Function}{: }{end}
\Fn{ \text{MPNN} $ \{e_{ki}, \mathbf{v}_i, c \} $}{
\For{m = 0, 1}{
\ForEach{\text{node} $i$, }{
\ForEach{$k \in \mathcal{N}(i)$ }{
$\mathbf{x}_{ki} = e_{ki} \cdot \mathbf{v}_k^m$ ; \\
		$\mathbf{x}_{ki} = \text{LReLU}(\mathbf{W}_m\mathbf{x}_{ki}+ \mathbf{b}_m) $ ;\\
		}
		$\mathbf{\bar{x}}_i = \frac{1}{|\mathcal{N}(i)|}\sum_{k\in \mathcal{N}(i)} \mathbf{x}_{ki}$ ; \\
		$\mathbf{v}_i^{m} = \text{LReLU} (\mathbf{W}_m\mathbf{v}_{i}^{m}+ \mathbf{b}_m) $ ; \\
		$\mathbf{v}_i^{m} = \mathbf{v}_i^{m} + \mathbf{\bar{x}}_i $; \\
		$\mathbf{v}_i^{m} = \text{Dropout}(\mathbf{v}_i^{m}) $
	}
}
$\mathbf{\bar{v}} = \frac{1}{M} \sum_{i} \mathbf{v}_i^1  $ ; \\
$\mathbf{\bar{v}} = \mathbf{W}_2\mathbf{\bar{v}} + \mathbf{b}_2 $ ; \\
$\hat{c} = \text{softmax} (\mathbf{\bar{v}})$ ; \\
return $ \hat{c}$ }
\caption{Message Passing Neural Network. 
}
\end{algorithm}
In Alg.~\ref{MPNN_algo}, LReLU stands for leaky ReLU activation and Dropout is the random dropout algorithm. The weight $\mathbf{W}_{m}$ and bias $\mathbf{b}_{m}$, where $m = 0,1,2$, are learnable parameters. Each node vector marked with $m=0,1$ are updated twice. In each updating, $\mathbf{W}_{m}$ and $\mathbf{b}_{m}$ are designed such that the length of each node vector gets longer and hence incorporates more information before the last global averaging. Besides, an affine function based on learned representation frees us from a direct update on the original edge attribute $e_{ki}$.  Finally, random dropout is used to battle over-fitting when training the network on the small dataset. 


\subsection{Chebyshev Convolution}

With vertex-wised signal $x$, the convolution operation on graph $\mathcal{G}$ such as $x * \mathcal{G}$ can also be defined in the Fourier domain. To analysis graph $\mathcal{G}$ in Fourier domain, one essential operation is to obtain the graph Laplacian $L=D-E$ where $E=\{\mathbf{e}_k\} = \{\mathbf{e}_{ij}\}$ and $D$ is the diagonal degree matrix with $D_{ii}=\sum_{j}\mathbf{e}_{ij}$  or as a normalized form $L=I_{n}-D^{(-1/2)} ED^{(-1/2)}$. The Laplacian can be diagonalized by its graph Fourier modes $U=[u_0,u_1,\dots,u_{M-1}]$ such that $L=U\Lambda U^T$ where $\Lambda=\mathrm{diag}\big([\lambda_0,\lambda_1, \dots,\lambda_{M-1} ]\big)$ are the frequencies of the graph. Finally, a signal $x$ is filtered by $g_\theta$ as
\begin{equation}
    y=g_\theta\big(L\big)x = g_\theta\big(U\Lambda U^T\big)x = Ug_\theta(\Lambda)U^Tx.
\end{equation}
Convolution with constant neighbors could use non-parametric filter such as $g_\theta (\Lambda)=diag(\theta)$ where $\theta \in \mathbb{R}^M$ is trainable variables. However, it is not localized in space. Therefore, polynomial parametrization is used to construct the localized filters $g_\theta (\Lambda)=\sum_{k=0}^{K-1}\theta_k \Lambda^k$ which limit the shortest path distance to $K$ (e.g. within the $K$-th order neighbors of a vertex)~\cite{defferrard2016convolutional}. In order to further accelerate the computation of the multiplication with the Fourier basis $U$, Chebyshev polynomial $T_k (x)$ is used to build the filter
\begin{equation}
    g_\theta \big(\Lambda \big) = \sum_{k=0}^{K-1}\theta_k T_k\big(\tilde{\Lambda} \big).
\end{equation}
As a result, the filtering operation can be written as 
\begin{equation}
 y=g_\theta \big(L\big)x=\sum_{k=0}^{K-1}\theta_k T_k\big(\tilde{L}\big)x   
\end{equation}
where $ \theta_k$ is the trainable parameters for a single output channel.

Pooling operation is another essential operation for building the convolution neural network (CNN), which gradually increases the receipt field size in each layer to enhance the hierarchical representation of features. The pooling operation on graphs is to cluster similar vertices depending on the meaningful neighborhoods. After clustering, all vertices within the same cluster will be represented as one vertex. We also require the clustering techniques that can reduce the size of the graph by a factor of two at each level, which offers a precise control of the coarsening and polling size. Since graph clustering is \textit{NP}-hard, approximation based algorithm are usually adapted in this calculation~\cite{fey2019fast}.

Beside the convolution and pooling layers, other operations such as activation, fully-connected, dropout and loss function are directly portable from the ordinary CNN defined on the rigid neighborhood.

\section{Experiments}

\subsection{Dataset}

We collected 530 half-Heusler compounds from the Materials Project database \cite{jain2013commentary} using the data mining approach. Each of the generated raw Hamiltonian sample contains three atoms, each with maximum 16 orbitals consisting of one $s$, three $p$, five $d$, and seven $f$ orbitals. The Hamiltonian matrices are all calculated within $7\times7\times7 = 343$ unit cells. 
The real values of target band gap fall in the range of $\big[0, 5.6\big]$. We choose a threshold 0.2 to produce binary labels ($E_g > 0.2$ for $ = 1$ and $E_g < 0.1$ as $c = 0$), which results 116 positive  and 414 negative samples. For balance, 117 negative samples are sampled from the original 414 ones to match the number of positive samples. Finally, a balanced subset with 233 samples are used in the experiments.

\subsection{Features}

We generate two groups of features for GCN based methods and shallow methods respectively. The feature used in GCN based approaches is already presented before.

For shallow methods, a set of fixed length features is created to include both atomic and interaction information. To limit the total dimension of feature, only the Hamiltonian from the center unit cell (relative coordinate $[0, 0, 0]$) is selected for use. Zeros padding is used to accommodate size variation of the Hamiltonian for different samples. In detail, all Hamiltonian are embedded in the square matrix of size $48\times48$ where each side corresponds to the three atoms each with 16 orbitals. The interaction features are the vectorization of the square matrix, which results features with a dimension of 2,304. The atomic features are the concatenation of the atomic number and atom coordinate feature in Tab.~\ref{tab:feat}. Finally, the combination of the two features results in a set of features with a dimension of $3 + 3\times3 + 2304 = 2316$.

\subsection{Experiment Settings}

Five popular shallow classification methods are evaluated, including Decision Tree, NaiveBayes, MultiLayer Perceptron, SVM and RandomForest. Three variations of the features are used in these experiments: (1) Interaction only, (2) Atom only,  and (3) Atom+Interaction. Since the class distribution is balanced, the baseline performance of random output is $50\%$. Reported results are binary classification accuracy by 5-fold cross-validation.

A two-layer MPNN and a three layer Chebyshev convolutional network are built upon the Hamiltonian data. For Chebyshev convolutional network, the convolutional filter size in those three layers is chosen as \{1, 2, 2\} respectively to gradually increase the receptive field. Leaky-ReLU is adapted for nonlinear activation. At last, a global average pooling layer followed by a softmax layer outputs the probabilities of input graph belong to the two categories. For both methods, Adam optimizer with a fixed learning rate of 0.001 and weight decay of 5e-4 is used. The training is performed up to 2,000 epochs. The two GCNs are implemented in PyTorch Geometric~\cite{Fey/Lenssen/2019} and trained on an Nvidia Titan X GPU.  

\subsection{Results}

\begin{figure}
	\includegraphics[width=\linewidth]{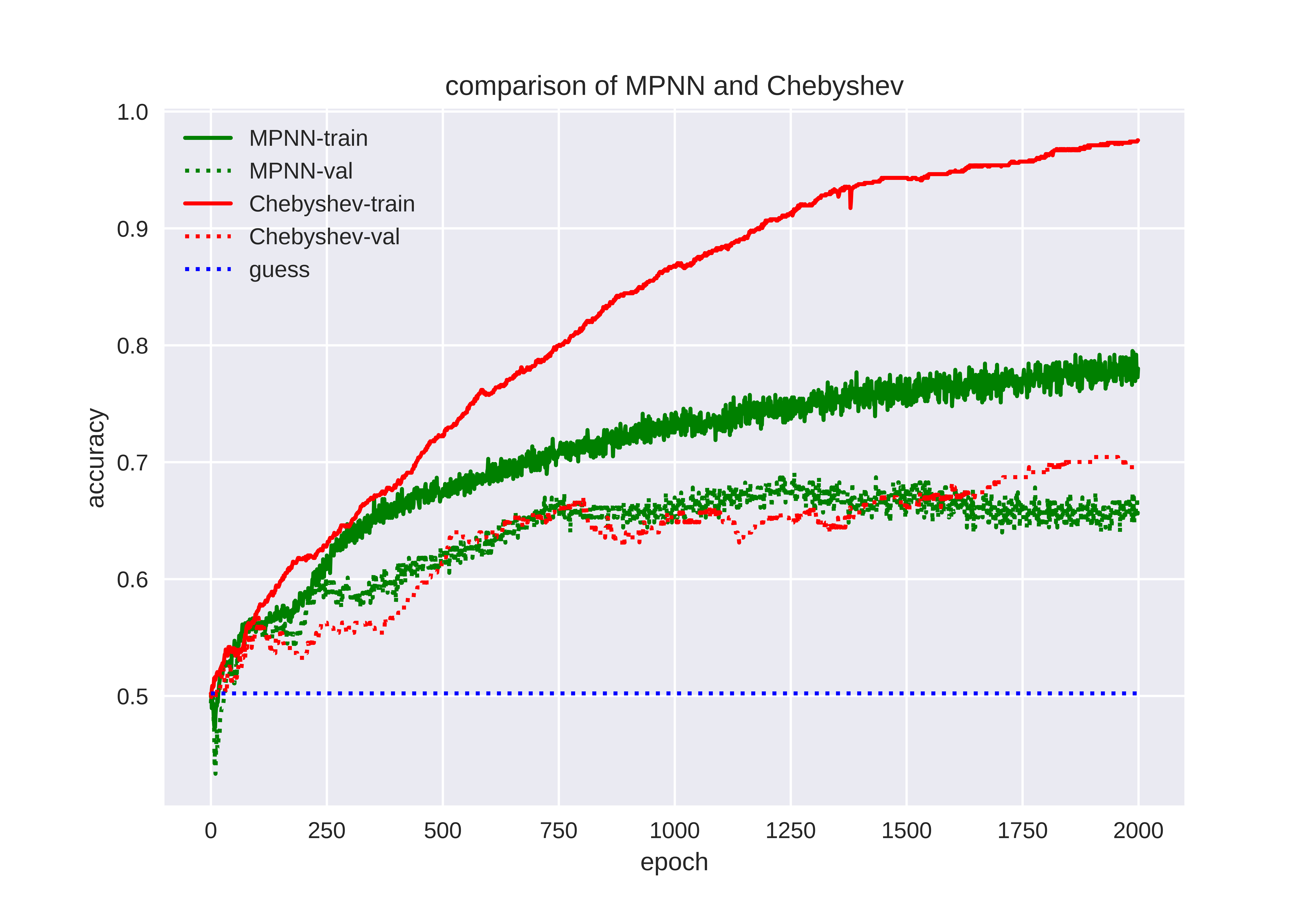}
	\caption{Learning curves of the two GCN based methods on one fold. }
	\label{accuracy_mine}
\end{figure}

\begin{table*}[]
	\caption{Results on the success rates (\%) of band gap classification.}
	\label{tab:res}
    \centering
    \begin{tabular}{c|c|c|c|c}
    \hline\hline
          & Method & Interaction & Atom & Atom+Interaction\\
         \hline
         \multirow{5}{*}{\rotatebox[origin=c]{90}{shallow}}  
         &Decision Tree         &   52.75 & 63.59 & 57.83 \\
         &NaiveBayes            &   57.03 & 57.45 & 61.78 \\
         &MultiLayer Perceptron &   64.91 & 52.84 & 66.16 \\
         &SVM                   &   67.89 & 55.40 & 66.23 \\
         &RandomForest          &   66.16 & 64.80 & 67.84 \\
         \hline
         \multirow{3}{*}{\rotatebox[origin=c]{90}{GCN}}  
         & MPNN w/o neighbor cell & - & - &69.12\\
         \cline{2-5}
         & Cheb. Conv. w/o neighbor cell & - & - &70.39\\
         & Cheb. Conv. w/ 1st neighbor cells & - & - & 70.43 \\
         
         \hline\hline
    \end{tabular}
\end{table*}




We report the classification accuracy of the shallow methods using the five-fold cross-validation in Tab.~\ref{tab:res}. Among the different variations of features, the Atom+Interaction usually achieves the best performance, while Atom feature alone performs worst. The Interaction features alone achieve the similar performance with the combination of Interaction and Atom feature, which demonstrates the importance of using the Hamiltonian to represent the whole characteristic of a material. The experiments on shallow methods confirm that the interaction embedded in the Hamiltonian contains the key information for the band gap prediction. Therefore, using graph to model this interaction structure of the Hamiltonian and applying GCN for learning process provide an advanced solution to this task.

The learning curves of MPNN and Chebyshev Network are shown in the Fig.~\ref{accuracy_mine}. The same cross-validation set generated with the same random seed are used here for fair comparison. Without the random dropout module, the Chebyshev Network clearly has over fitting on the validation accuracy curve with nearly $68\%$ accuracy while it is $98\%$ accuracy on the training set. By applying the random dropout module, there is only $10\%$ difference of accuracy in training and validation for MPNN. We also report numerical performance of all five folds in Tab.~\ref{tab:res}.

We also explore the case that includes the neighboring unit cell for the classification task. When only center unit cell was used, one can observe from the results that the GCN based method is obviously better than traditional shallow methods benefited by the more accurate and compact graph representation of the Hamiltonian data. By further including the neighboring unit cells, feature dimension will dramatically increases to 20,000 if following the same strategy mentioned above to unify the size of the input Hamiltonian for shallow methods. This high dimension input is unacceptable for shallow methods before effective dimension reduction technique is applied. But the GCN based methods can handle this case naturally. As shown in Tab.~\ref{tab:res}, Chebyshev convolution based network achieves similar performance with the eight times larger input. With the limited training samples, the high dimensional input may not directly benefit the final performance. When training with sufficient data and including other essential layers, such as local pooling and batch normalization layers, we can expect the GCN to achieve much better performance in learning the periodic structure information in the Hamiltonian for the physical properties prediction.

\section{Conclusion}

In this paper, we investigate two different GCN algorithms on the Hamiltonian data for band gap classification. The five-fold cross-validation results show that using graph based representation and learning techniques for the Hamiltonian data achieve advanced performance over other shallow methods combining hand-crafted dense features. Moreover, compared with traditional methods that require fixed input dimension, graph formulates the interaction structure in the Hamiltonian more naturally and compactly, which is scalable to include the neighboring cells for exploring the periodic pattern in the condensed matter system.

\textbf{Acknowledgement:} N.W. and X.Q. gratefully acknowledge the support by the National Science Foundation (NSF) under award number OAC-1835690. Portions of this research were conducted with the advanced computing resources provided by Texas A\&M High Performance Research Computing. Q.Y. acknowledges the support by the U.S. Department of Energy, Office of Science, under award number DE-SC0020310.

{\small
\bibliographystyle{plain}
\bibliography{egbib}

\begin{thebibliography}{10}

\bibitem{mostofi2014wannier}
An updated version of wannier90: A tool for obtaining maximally-localised
  wannier functions.
\newblock {\em Computer Physics Communications}, 185(8):2309 -- 2310, 2014.

\bibitem{Bartok2010}
A.~P. Bartok, M.~C. Payne, R.~Kondor, and G.~Csanyi.
\newblock Gaussian approximation potentials: The accuracy of quantum mechanics,
  without the electrons.
\newblock {\em Physical Review Letters}, 104(13):136403, 2010.

\bibitem{battaglia2018relational}
Peter~W Battaglia, Jessica~B Hamrick, Victor Bapst, Alvaro Sanchez-Gonzalez,
  Vinicius Zambaldi, Mateusz Malinowski, Andrea Tacchetti, David Raposo, Adam
  Santoro, Ryan Faulkner, et~al.
\newblock Relational inductive biases, deep learning, and graph networks.
\newblock {\em arXiv preprint arXiv:1806.01261}, 2018.

\bibitem{Behler2011}
J.~Behler.
\newblock Neural network potential-energy surfaces in chemistry: a tool for
  large-scale simulations.
\newblock {\em Physical Chemistry Chemical Physics}, 13(40):17930--17955, 2011.

\bibitem{bronstein2017geometric}
Michael~M Bronstein, Joan Bruna, Yann LeCun, Arthur Szlam, and Pierre
  Vandergheynst.
\newblock Geometric deep learning: going beyond euclidean data.
\newblock {\em IEEE Signal Processing Magazine}, 34(4):18--42, 2017.

\bibitem{Chandrasekaran2019}
Anand Chandrasekaran, Deepak Kamal, Rohit Batra, Chiho Kim, Lihua Chen, and
  Rampi Ramprasad.
\newblock Solving the electronic structure problem with machine learning.
\newblock {\em npj Computational Materials}, 5(1):22, 2019.

\bibitem{Curtarolo2003}
S.~Curtarolo, D.~Morgan, K.~Persson, J.~Rodgers, and G.~Ceder.
\newblock Predicting crystal structures with data mining of quantum
  calculations.
\newblock {\em Physical Review Letters}, 91(13):135503, 2003.

\bibitem{defferrard2016convolutional}
Micha{\"e}l Defferrard, Xavier Bresson, and Pierre Vandergheynst.
\newblock Convolutional neural networks on graphs with fast localized spectral
  filtering.
\newblock In {\em Advances in neural information processing systems}, pages
  3844--3852, 2016.

\bibitem{Dong2019}
Yuan Dong, Chuhan Wu, Chi Zhang, Yingda Liu, Jianlin Cheng, and Jian Lin.
\newblock Bandgap prediction by deep learning in configurationally hybridized
  graphene and boron nitride.
\newblock {\em npj Computational Materials}, 5(1):26, 2019.

\bibitem{dwivedi2020benchmarking}
Vijay~Prakash Dwivedi, Chaitanya~K Joshi, Thomas Laurent, Yoshua Bengio, and
  Xavier Bresson.
\newblock Benchmarking graph neural networks.
\newblock {\em arXiv preprint arXiv:2003.00982}, 2020.

\bibitem{Fey/Lenssen/2019}
Matthias Fey and Jan~E. Lenssen.
\newblock Fast graph representation learning with {PyTorch Geometric}.
\newblock In {\em ICLR Workshop on Representation Learning on Graphs and
  Manifolds}, 2019.

\bibitem{fey2019fast}
Matthias Fey and Jan~Eric Lenssen.
\newblock Fast graph representation learning with pytorch geometric.
\newblock {\em arXiv preprint arXiv:1903.02428}, 2019.

\bibitem{Ghiringhelli2015}
L.~M. Ghiringhelli, J.~Vybiral, S.~V. Levchenko, C.~Draxl, and M.~Scheffler.
\newblock Big data of materials science: Critical role of the descriptor.
\newblock {\em Physical Review Letters}, 114(10):105503, 2015.

\bibitem{gilmer2017neural}
Justin Gilmer, Samuel~S Schoenholz, Patrick~F Riley, Oriol Vinyals, and
  George~E Dahl.
\newblock Neural message passing for quantum chemistry.
\newblock In {\em Proceedings of the 34th International Conference on Machine
  Learning-Volume 70}, pages 1263--1272. JMLR. org, 2017.

\bibitem{Hansen2013}
Katja Hansen, Grégoire Montavon, Franziska Biegler, Siamac Fazli, Matthias
  Rupp, Matthias Scheffler, O.~Anatole von Lilienfeld, Alexandre Tkatchenko,
  and Klaus-Robert Müller.
\newblock Assessment and validation of machine learning methods for predicting
  molecular atomization energies.
\newblock {\em Journal of Chemical Theory and Computation}, 9(8):3404--3419,
  2013.

\bibitem{Hautier2010}
G.~Hautier, C.~C. Fischer, A.~Jain, T.~Mueller, and G.~Ceder.
\newblock Finding nature's missing ternary oxide compounds using machine
  learning and density functional theory.
\newblock {\em Chemistry of Materials}, 22(12):3762--3767, 2010.

\bibitem{henaff2015deep}
Mikael Henaff, Joan Bruna, and Yann LeCun.
\newblock Deep convolutional networks on graph-structured data.
\newblock {\em arXiv preprint arXiv:1506.05163}, 2015.

\bibitem{jain2013commentary}
Anubhav Jain, Shyue~Ping Ong, Geoffroy Hautier, Wei Chen, William~Davidson
  Richards, Stephen Dacek, Shreyas Cholia, Dan Gunter, David Skinner, Gerbrand
  Ceder, et~al.
\newblock Commentary: The materials project: A materials genome approach to
  accelerating materials innovation.
\newblock {\em APL Materials}, 1(1):011002, 2013.

\bibitem{kipf2016semi}
Thomas~N Kipf and Max Welling.
\newblock Semi-supervised classification with graph convolutional networks.
\newblock 2017.

\bibitem{krizhevsky2012imagenet}
Alex Krizhevsky, Ilya Sutskever, and Geoffrey~E Hinton.
\newblock Imagenet classification with deep convolutional neural networks.
\newblock In {\em Advances in neural information processing systems}, pages
  1097--1105, 2012.

\bibitem{lecun1998gradient}
Yann LeCun, L{\'e}on Bottou, Yoshua Bengio, and Patrick Haffner.
\newblock Gradient-based learning applied to document recognition.
\newblock {\em Proceedings of the IEEE}, 86(11):2278--2324, 1998.

\bibitem{Lee2016}
Joohwi Lee, Atsuto Seko, Kazuki Shitara, Keita Nakayama, and Isao Tanaka.
\newblock Prediction model of band gap for inorganic compounds by combination
  of density functional theory calculations and machine learning techniques.
\newblock {\em Physical Review B}, 93(11):115104, 2016.

\bibitem{Long2007}
C.~J. Long, J.~Hattrick-Simpers, M.~Murakami, R.~C. Srivastava, I.~Takeuchi,
  V.~L. Karen, and X.~Li.
\newblock Rapid structural mapping of ternary metallic alloy systems using the
  combinatorial approach and cluster analysis.
\newblock {\em Review of Scientific Instruments}, 78(7):072217, 2007.

\bibitem{marzari1997mlwf}
Nicola Marzari and David Vanderbilt.
\newblock Maximally localized generalized wannier functions for composite
  energy bands.
\newblock {\em Physical Review B}, 56:12847--12865, 1997.

\bibitem{Pilania2016}
G.~Pilania, A.~Mannodi-Kanakkithodi, B.~P. Uberuaga, R.~Ramprasad, J.~E.
  Gubernatis, and T.~Lookman.
\newblock Machine learning bandgaps of double perovskites.
\newblock {\em Scientific Reports}, 6:19375, 2016.

\bibitem{qian2008qo}
Xiaofeng Qian, Ju~Li, Liang Qi, Cai-Zhuang Wang, Tzu-Liang Chan, Yong-Xin Yao,
  Kai-Ming Ho, and Sidney Yip.
\newblock Quasiatomic orbitals for ab initio tight-binding analysis.
\newblock {\em Physical Review B}, 78:245112, 2008.

\bibitem{Rajan2005}
K.~Rajan.
\newblock Materials informatics.
\newblock {\em Materials Today}, 8(10):38--45, 2005.

\bibitem{Shi2019}
Zhe Shi, Evgenii Tsymbalov, Ming Dao, Subra Suresh, Alexander Shapeev, and
  Ju~Li.
\newblock Deep elastic strain engineering of bandgap through machine learning.
\newblock {\em Proceedings of the National Academy of Sciences}, 116(10):4117,
  2019.

\bibitem{Snyder2012}
J.~C. Snyder, M.~Rupp, K.~Hansen, K.~R. Muller, and K.~Burke.
\newblock Finding density functionals with machine learning.
\newblock {\em Physical Review Letters}, 108(25):253002, 2012.

\bibitem{zhou2018graph}
Jie Zhou, Ganqu Cui, Zhengyan Zhang, Cheng Yang, Zhiyuan Liu, and Maosong Sun.
\newblock Graph neural networks: A review of methods and applications.
\newblock {\em arXiv preprint arXiv:1812.08434}, 2018.

\end{thebibliography}
}

\end{document}